\documentclass[aps,twocolumn,showpacs]{revtex4}
\usepackage{color}
\usepackage{graphicx}
\usepackage{graphics}
\usepackage{epsfig}
\usepackage{subfigure}
\usepackage{latexsym}
\usepackage{amsfonts}
\usepackage[english]{babel}
\usepackage[latin1]{inputenc}
\usepackage[all]{xy}
\usepackage{amsmath}
\usepackage{amssymb}
\newcommand{\be}{\begin{equation}}
\newcommand{\ee}{\end{equation}}
\newcommand{\ben}{\begin{eqnarray}}
\newcommand{\een}{\end{eqnarray}}
\newcommand{\bes}{\begin{subequations}}
\newcommand{\ees}{\end{subequations}}

\def \su{\mathop{\rm su}\nolimits}
\def \sv{\mathop{\rm sv}\nolimits}
\def \sf{\mathop{\rm sf}\nolimits}

\def \cu{\mathop{\rm cu}\nolimits}
\def \cv{\mathop{\rm cv}\nolimits}
\def \cf{\mathop{\rm cf}\nolimits}

\newcommand{\ba}[1]{\begin{array}{#1}}
\newcommand{\ea}{\end{array}}
\newcommand{\bea}[1]{\begin{equation}\left\{\begin{array}{#1}}
\newcommand{\eea}{\end{array}\right.\end{equation}}

\begin{document}
\title{Kinks in a non-linear massive sigma model}
\author{A. Alonso-Izquierdo$^{a}$, M. A. Gonzalez Leon$^{a}$ and J. Mateos Guilarte$^{b}$}
\affiliation{{$^{a}$ Departamento de Matematica Aplicada and IUFFyM,
Universidad de Salamanca, SPAIN}
\\{$^{b}$Departamento de Fisica and IUFFyM, Universidad de Salamanca, SPAIN}}

\begin{abstract}
We describe the kink solitary waves of a massive non-linear sigma
model with an ${\mathbb S}^2$ sphere as the target manifold. Our
solutions form a moduli space of non-relativistic solitary waves in
the long wavelength limit of ferromagnetic linear spin chains.
\end{abstract}
 \maketitle


\section{ The massive non-linear ${\mathbb S}^2$-sigma model}

In this letter we shall concentrate on the 1D non-linear ${\mathbb
S}^2$-sigma model: (a) The space-time is the (1+1)-dimensional
${\mathbb R}^{1,1}$ Minkowski space. $x^\mu,\, \mu=0,1$ ($ x^0\equiv
t, x^1\equiv x$) denotes a point in ${\mathbb R}^{1,1}$. We choose
the metric in the form $g^{\mu\nu}={\rm diag}(1,-1)$ and the
d'Alembertian reads: $\Box=\frac{\partial^2}{\partial
t^2}-\frac{\partial^2}{\partial x^2}$. (b) The target (internal)
space is the ${\mathbb S}^2$-sphere. This is in contrast with the
original Gell-Mann/L{\`e}vy model where the target space is
${\mathbb S}^3$ \cite{Gell-Mann}. Three scalar fields
$\Phi=(\phi_a)$, $a=1,2,3$, define a map $ \Phi :{\mathbb R}^{1,1}
\to  {\mathbb S}^2$ if they are constrained to the surface in
${\mathbb R}^3$:
\begin{equation}
\phi_1^2(x^\mu)+\phi_2^2(x^\mu)+\phi_3^2(x^\mu)=R^2\label{sphere}
\end{equation}

\noindent The action
\begin{equation}
S=\int \, dtdx \,
\left\{\frac{1}{2}g^{\mu\nu}\sum_{a=1}^3\frac{\partial\phi_a}{\partial
x^\mu}\frac{\partial\phi_a}{\partial x^\nu}\right\}\label{action2}
\end{equation}
seems to be simple but together with the constraint (\ref{sphere})
governs the complicated non-linear dynamics of two Goldstone bosons
with coupling constant $\frac{1}{R}$. In one spatial dimension,
however, Goldstone particles do not exist \cite{Col}. The infrared
asymptotics of the (\ref{sphere})-(\ref{action2}) system induces a
potential energy density that we choose as:
\begin{equation}
V(\phi_1,\phi_2)=\frac{\lambda^2}{2}\phi_1^2(t,x)+\frac{\gamma^2}{2}\phi_2^2(t,x)\label{GL-pot}
\end{equation}
giving masses $\lambda$ and $\gamma$ to the massless excitations.

With no loss of generality, we assume that: $\lambda^2\geq\gamma^2
>0$, and we define the non-dimensional parameter: $
\sigma^2=\frac{\gamma^2}{\lambda^2}$, $0<\sigma^2 \leq 1$, measuring
the ratio between particle masses. We also re-scale the space-time
coordinates to address non-dimensional variables: $ x^\mu \to
\frac{x^\mu}{\lambda}$. Solving the constraint the action reads
\begin{eqnarray}
S&=&\frac{1}{2}\int \, dtdx \, \left\{\sum_{\alpha=1}^2\,
\frac{\partial\phi_\alpha}{\partial
x^\mu}\frac{\partial\phi_\alpha}{\partial
x_\mu}-\phi_1^2-\sigma^2\phi_2^2\right.\nonumber
\\&+&\left.
\frac{\sum_{\alpha=1}^2(\phi_\alpha\partial^\mu\phi_\alpha)
\sum_{\beta=1}^2(\phi_\beta\partial_\mu\phi_\beta)}{R^2-\phi_1^2-\phi_2^2}\right\}
\label{action3}
\end{eqnarray}

Despite the potential energy density being quadratic, there are two
homogeneous minima of the action (vacua): the North and South Poles:
$ \Phi^A=(0,0,R) , \Phi^{\bar{A}}= (0,0,-R)$. Thus, the discrete
symmetry of the action(\ref{action3}) $ {\mathbb Z}_2\times{\mathbb
Z}_2\times{\mathbb Z}_2 $ generated by $\phi_a \rightarrow -\phi_a$,
$a=1,2,3$ is spontaneously broken to $ {\mathbb Z}_2\times{\mathbb
Z}_2 $ (generated by $ \phi_\alpha \rightarrow -\phi_\alpha$,
$\alpha=1,2$). If the two masses were equal, this unbroken symmetry
would become the $SO(2)$ rotation group around the North-South Pole
axis.

\section{Topological Kinks}

Using spherical coordinates, $ \phi_1=R\cos\varphi\sin\theta$, $
\phi_2=R\sin\varphi\sin\theta$, $\phi_3=R\cos\theta$, in the chart
${\mathbb S}^2-\{\bar{A}\}$ of ${\mathbb S}^2$ the energy of static
configurations: $\theta(t,x)=\theta(x)\in[0,\pi)$,
$\varphi(t,x)=\varphi(x)\in[0,2\pi)$, $E=\int dx\, {\cal
E}(\theta(x),\varphi(x))$, and the potential energy density read:
\[
E=\int dx
\left\{\frac{R^2}{2}\left[\left(\frac{d\theta}{dx}\right)^2+\sin^2\theta\left(
\frac{d\varphi}{dx}\right)^2\right]+V(\theta,\varphi)\right\}
\]
\begin{equation}
V(\theta(x),\varphi(x)) = \frac{R^2}{2} \sin^2 \theta(x) (\sigma^2 +
(1-\sigma^2) \cos^2\varphi(x))\label{GL-pot-polar}
\end{equation}

The configuration space of the system ${\cal C}=\left\{{\rm
Maps}({\mathbb R},{\mathbb S}^2)/E<+\infty \right\}$ is formed by
four disconnected sectors according to the tendency of every finite
energy configuration towards either the North or South Pole at the
extremes of the spatial line $x=\pm\infty$.

Solutions for which the temporal dependence is of the form
\[
\theta(t,x)=\theta\left(\frac{x-vt}{\sqrt{1-v^2}}\right) \, ,
\varphi(t,x)=\varphi\left(\frac{x-vt}{\sqrt{1-v^2}}\right)
\]
for some velocity $v$, are solitary or traveling waves. Lorentz
invariance provides all the solitary waves from the static solutions
of the field equations
\begin{eqnarray}
\frac{d^2\theta}{dx^2}-\frac{\sin
2\theta}{2}\left(\frac{d\varphi}{dx}\right)^2 &=& \frac{\sin
2\theta}{2}
\left(\cos^2\varphi+\sigma^2\sin^2\varphi\right)\label{field-eq1}\\
\frac{d}{d x}\left(\sin^2\theta\frac{d\varphi}{d
x}\right)&=&\frac{1-\sigma^2}{2}\sin^2\theta\sin
2\varphi\label{field-eq2}
\end{eqnarray}

On the orbits $\varphi_{K_1}(x)=\pm\frac{\pi}{2}$ (half meridians)
the system (\ref{field-eq1})-(\ref{field-eq2}) becomes the ODE of
the pendulum and the separatrix trajectories between bounded and
unbounded motion;
\[
-\frac{d^2\theta}{dx^2}+\frac{\sigma^2}{2}\sin 2\theta=0\Rightarrow
\theta_{K_1}(x)=2\arctan e^{\pm\sigma(x-x_0)}
\]
are the solitary waves or kinks (at their center of mass) of finite
energy: $E_{K_1}=2 R^2 \sigma$, interpolating between the North $A$
and South $\bar{A}$ Poles. Thus, these kinks belong to a different
sector in configuration space than the vacua, an evident fact in
Cartesian coordinates; see Figures 1-2 \, \footnote{In the web page
{\it http://www.usal.es/$\sim$mpg/} (Mathematica tools) there is a
Mathematica file containing several extra figures and formulas
concerning this work.}:
\[
\Phi^{K_1}(x)=\left( 0, \frac{1}{{\rm cosh}[\sigma(x-x_0)]},\ \pm
{\rm tanh}[\sigma(x-x_0)]\right)
\]

\begin{figure}[htbp]
\centerline{\includegraphics[height=2.5cm]{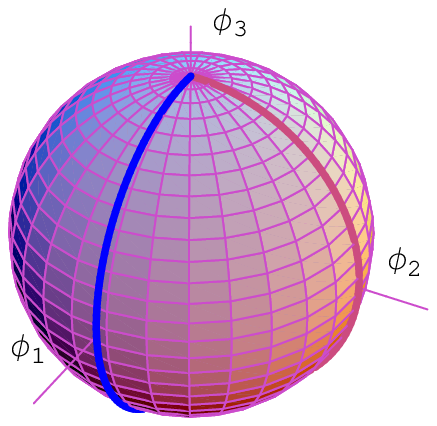}\quad
\includegraphics[height=2.5cm]{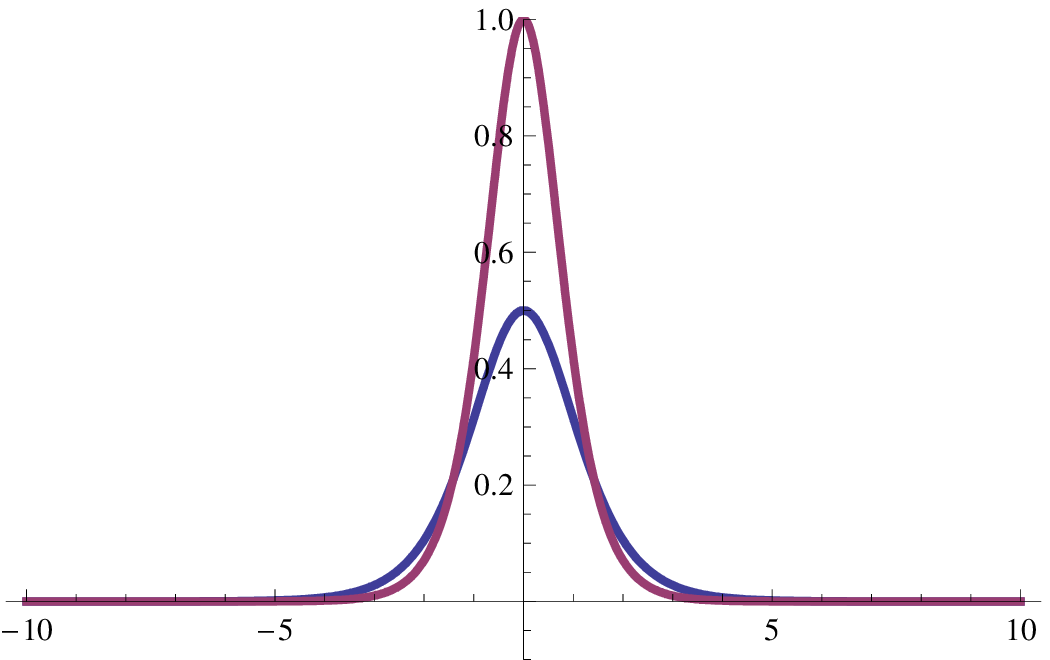}} \caption{a) $K_1$ and $K_2$
($\sigma^2=\frac{1}{2}$) kink orbits. b) $K_1$ (blue) and $K_2$
(red) kink energy densities}
\end{figure}

\begin{figure}[htbp]
\centerline{\includegraphics[height=2.5cm]{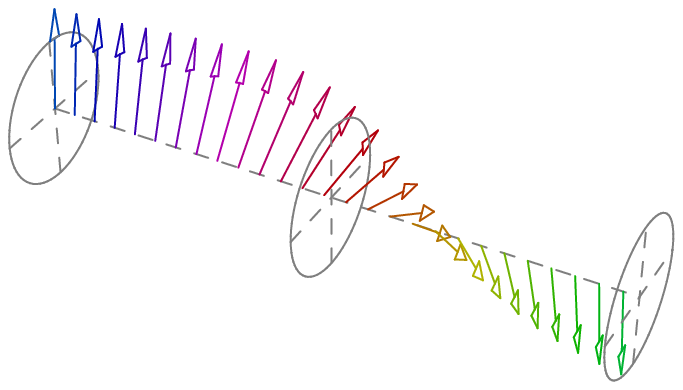}\quad
\includegraphics[height=2.5cm]{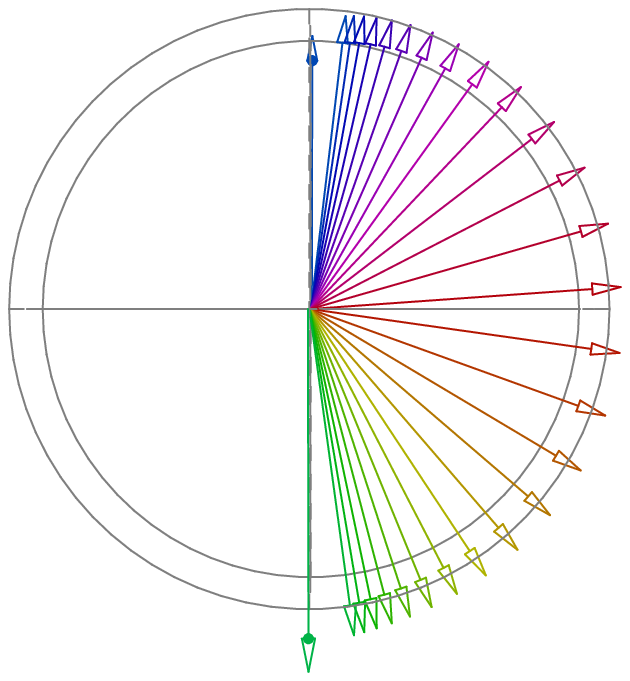}
} \caption{a) $K_1$ kinks. b) Perspective from one component of the
boundary of the target sphere $\times$ the spatial line infinite
cylinder: ${\mathbb S}^2\times{\mathbb R}$. The meridian
$\varphi_{K_1}=\frac{\pi}{2}$, $\varphi_{K_1}=-\frac{\pi}{2}$ is
plotted as orthogonal to the spatial line.}
\end{figure}

The half-meridians $\varphi_{K_2}(x)=0$ or $\varphi_{K_2}(x)=\pi$
are also good trial orbits in the sense that they provide new kinks
of finite energy $E_{K_2}=2 R^2$
\[
-\frac{d^2\theta}{dx^2}+\frac{1}{2}{\rm sin}2\theta=0\Rightarrow
\theta_{K_2}(x)=2\arctan e^{\pm(x-x_0)}
\]
via finite action solutions of another pendulum equation. These
solitary waves live in the same sector of the configuration space as
the previous ones and look similar in Cartesian coordinates; see
Figures 1-3:
\[
\Phi^{K_2}(x)=\left( \frac{1}{{\rm cosh}(x-x_0)},0,\pm {\rm
tanh}(x-x_0)\right)
\]

\begin{figure}[htbp]
\centerline{\includegraphics[height=2.5cm]{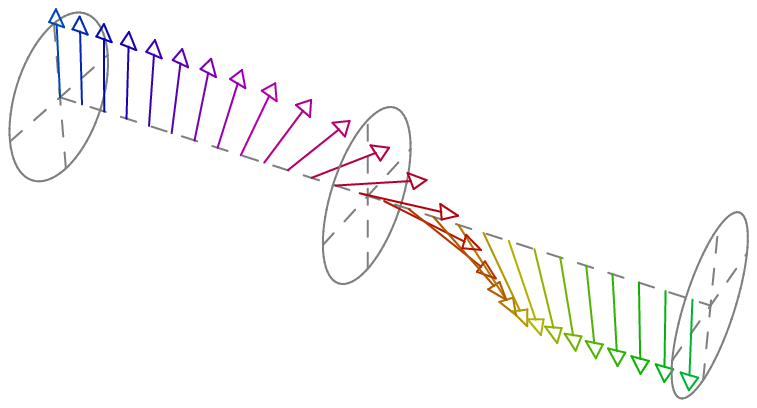}\quad
\includegraphics[height=2.5cm]{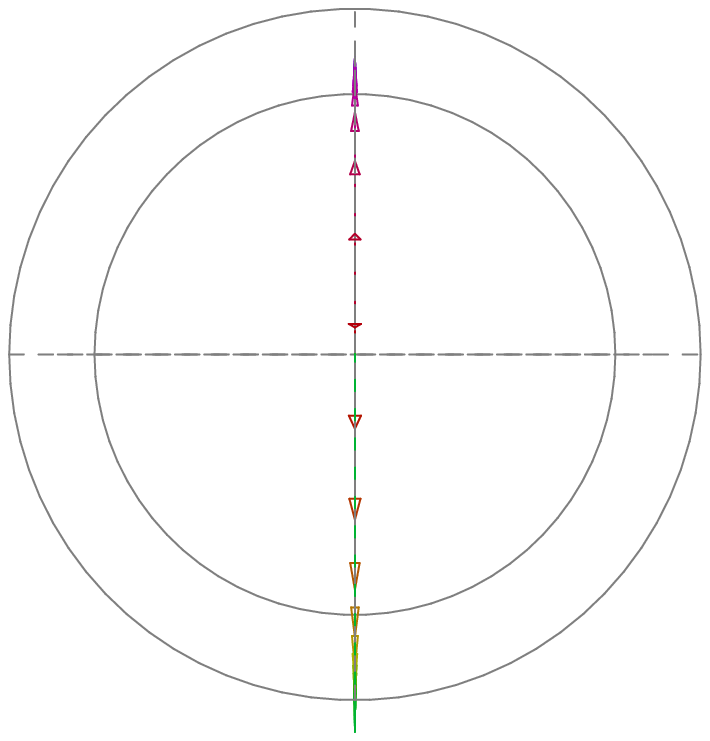}} \caption{a) $K_2$ kinks. b) Perspective
from one component of the boundary of the infinite cylinder:
${\mathbb S}^2\times{\mathbb R}$. The meridian $\varphi_{K_2}=0$,
$\varphi_{K_2}=\pi$, is plotted aligned with the spatial line.}
\end{figure}

\noindent At the $\sigma=1$ limit, all the half meridians
$\phi_{K_\alpha}(x)=\varphi \in [0, 2\pi)$ are good trial orbits and
there is a one-parametric family of solitary waves: $
\theta_{K_\varphi}(x)=2\arctan e^{\pm(x-x_0)}$,
\[
\Phi^{K_\varphi}(x)=\left( \frac{\cos\varphi}{\cosh (x-x_0)},
\frac{\sin \varphi}{\cosh (x-x_0)}, \pm \tanh(x-x_0)\right)
\]
degenerated in energy: $E_{K_\varphi}=2 R^2$, $\forall\varphi$.

\section{The non-linear massive sigma model in elliptic coordinates on the sphere}

We could now try to search for more kinks even in the case
$\sigma^2<1$ of distinct masses by using Rajaraman's trial orbit
method \cite{Raj}, but instead we shall profit from the fact that
the ODE system (\ref{field-eq1})-(\ref{field-eq2}) is integrable
using elliptic coordinates in the ${\mathbb S}^2$ sphere. In
${\mathbb S}^2$ we fix the two points: $ F_1\equiv (\theta_f,\pi)$,
$F_2\equiv (\theta_f,0)$, $\theta_f \in [0,\frac{\pi}{2})$. The
distance between them is: $d=2f=2R\theta_f<\pi R$, see Figure 4.

\begin{figure}[htbp]
\centerline{\includegraphics[height=2.5cm]{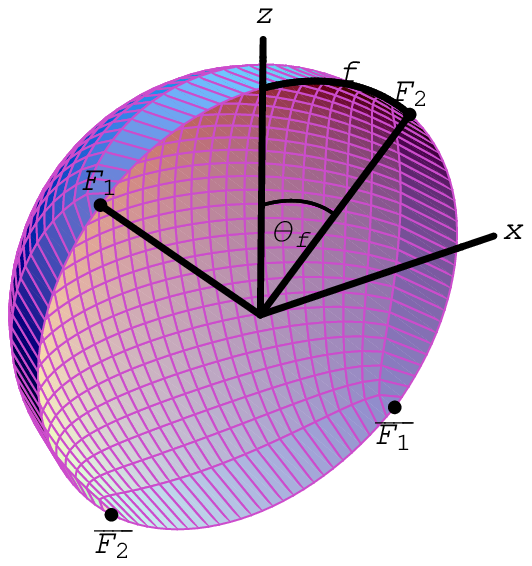}\quad
\includegraphics[height=2.5cm]{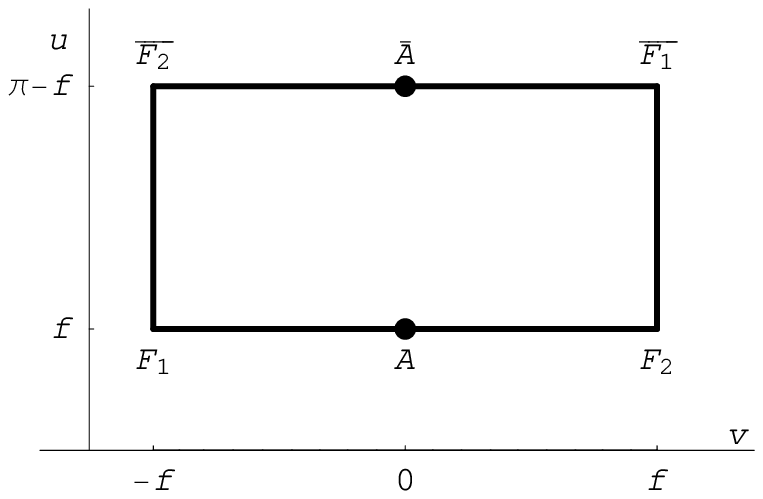}} \caption{Foci and antipodal foci of the
elliptic system of coordinates on the sphere.}
\end{figure}

\noindent Given a point $\Phi\in{\mathbb S}^2$, let us consider the
distances $r_1\in [0,\pi R]$ and $r_2\in [0,\pi R]$ from $\Phi$ to
$F_1$ and $F_2$. The elliptic coordinates of $\Phi$ are half the sum
and half the difference of $r_1$ and $r_2$: $\Phi\equiv
(u=\frac{r_1+r_2}{2}, v=\frac{r_1-r_2}{2})$. The formulae
\begin{eqnarray*}
&&\phi_1(t,x)=\frac{R}{\sf} \su \sv\, , \
\phi_2(t,x)=\frac{R}{\cf}\cu \cv\\&& \phi_3(t,x)=\pm
R\sqrt{1-\frac{\su^2 \sv^2}{\sf^2}-\frac{\cu^2 \cv^2}{\cf^2}}
\end{eqnarray*}
allow one to pass from elliptic to Cartesian coordinates. Here, a
simplified notation is used: $\su=\sin\frac{u(t,x)}{R}$,
$\cu=\cos\frac{u(t,x)}{R}$, $\sv=\sin\frac{v(t,x)}{R}$,
$\sf=\sin\theta_f$, etc. The differential arc-length in ${\mathbb
S}^2$ in this system of coordinates is 
\[
ds^2_{{\mathbb
S}^2}=\frac{1}{2}\cdot\frac{\su^2-\sv^2}{\su^2-\sf^2}\cdot
du^2+\frac{1}{2}\cdot\frac{\su^2-\sv^2}{\sf^2-\sv^2}\cdot dv^2
\]
Choosing the foci in such a way that $\cf^2=\sigma^2$, the energy
density of our system in elliptic coordinates reads:
\begin{eqnarray}
{\cal
E}[u,v]&=&\frac{1}{2}\left[\frac{\su^2-\sv^2}{\su^2-\sf^2}\left(\frac{du}{dx}\right)^2+
\frac{\su^2-\sv^2}{\sf^2-\sv^2}\left(\frac{dv}{dx}\right)^2\right]
\nonumber
\\ &&-\frac{f(u)+g(v)}{\su^2-\sv^2} \label{separation}
\end{eqnarray}
$f(u(x))=\frac{R^2}{2}\su^2(\su^2-\sf^2)$, $
g(v(x))=\frac{R^2}{2}\sv^2(\sf^2-\sv^2)$.

The mechanical analogy demands that we think of ${\cal E}$ as the
Lagrangian, $x$ as the time, $U[u(x),v(x)]=-V[u(x),v(x)]$ as the
mechanical potential energy, and the target manifold ${\mathbb S}^2$
as the configuration space. The structure of ${\cal E}[u,v]$ is such
that we are dealing with a Type I Liouville model \cite{Per} on the
sphere, i.e., a dynamical system which is Hamilton-Jacobi separable
in elliptic coordinates. The kink orbits (finite action
trajectories) and the kink profiles (\lq\lq time" schedules of these
trajectories) are given in the Hamilton-Jacobi framework
\cite{Ito}-\cite{AAi} via the quadratures: ($p_u=\frac{\partial
{\cal E}}{\partial \dot{u}}$, $p_v=\frac{\partial {\cal E}}{\partial
\dot{v}}$)
\[
 \int \frac{{\rm sg}\left(p_u\right)\, du}{\sqrt{2(\su^2-\sf^2)f(u)}}
 - \int \frac{{\rm sg}\left(p_v\right) dv}{\sqrt{2(\sf^2-\sv^2)g(v)}} = R^2 \gamma_2
\]
\[
\int \frac{{\rm sg}\left(p_u\right)\, \su^2
du}{\sqrt{2(\su^2-\sf^2)f(u)}} - \int \frac{{\rm
sg}\left(p_v\right)\, \sv^2 dv}{\sqrt{2(\sf^2-\sv^2)g(v)}} = x+
\gamma_1
\]

\section{Non-topological kinks}

In this way we find a family of non-topological kink (NTK) orbits by
integrating the first quadrature (they start and end either at the
North or the South Pole, they live in the vacua sectors of the
configuration space) parametrized by the integration constant
$C=e^{R^2 \gamma_2 \sf^2}$:
\[
C = \left[ \frac{\left|\tan \frac{u-f}{2R} \tan
\frac{u+f}{2R}\right|^{\frac{1}{2\cf }}}{|\tan \frac{u}{2R}|}
\right]^{{\rm sg}p_u} \left[ \frac{|\tan \frac{v}{2R}|}{\left|\tan
\frac{v-f}{2R} \tan \frac{v+f}{2R}\right|^{\frac{1}{2\cf }}}
\right]^{{\rm sg}p_v}
\]
The kink profiles of these non-topological solitary waves given by
the integration of the second quadrature
\[
e^{2(x+\gamma_1) \cf} =  \frac{\left|\tan \frac{u-f}{2R} \tan
\frac{u+f}{2R}\right|^{{\rm sg}p_u}}{\left|\tan \frac{v-f}{2R} \tan
\frac{v+f}{2R}\right|^{{\rm sg}p_v}}
\]
depend on one integration constant, $\gamma_1$, which sets the
center of each kink, see Figures 5 and 6.
\begin{figure}[htbp]
\centerline{\includegraphics[height=2.5cm]{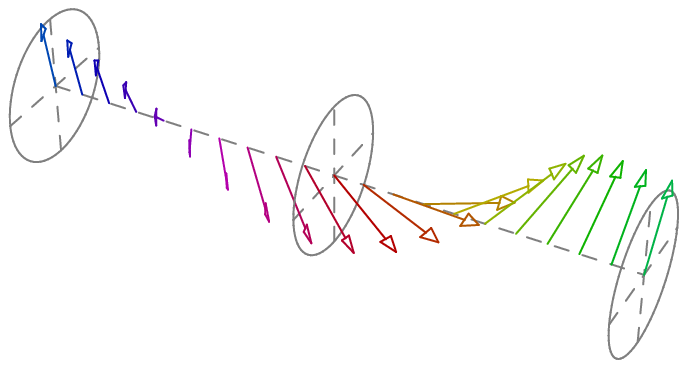}\qquad
\includegraphics[height=2.5cm]{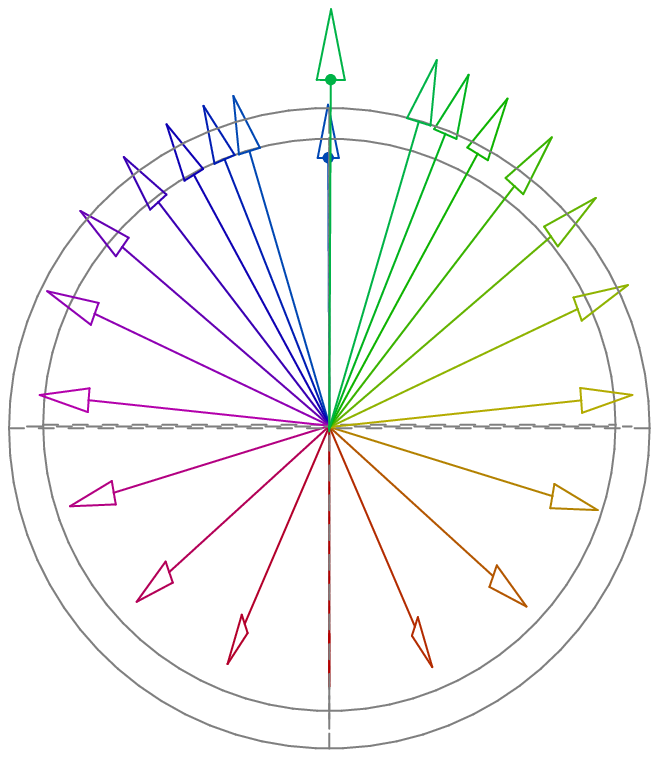}}
\caption{a) A $K_{\gamma_2} (|\gamma_2|<\infty)$ NTK kink. b)
Perpendicular cross section of the field variation.}
\end{figure}

\begin{figure}[htbp]
\centerline{\includegraphics[height=2.5cm]{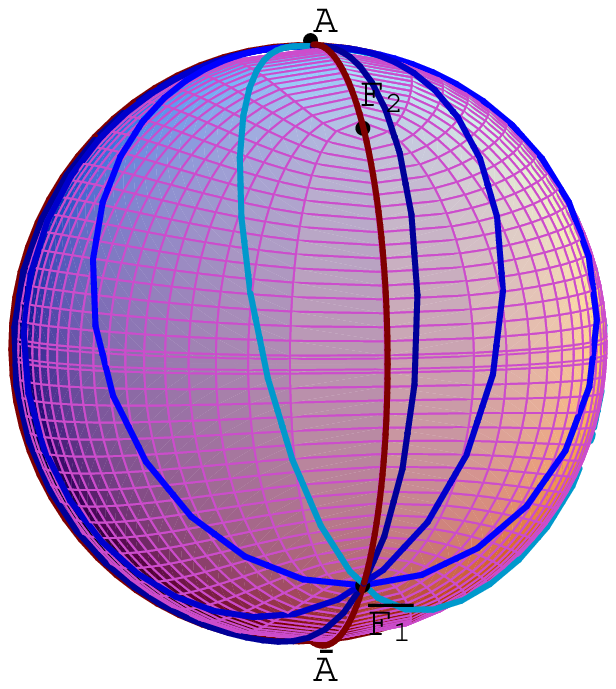}\qquad
\includegraphics[height=2.5cm]{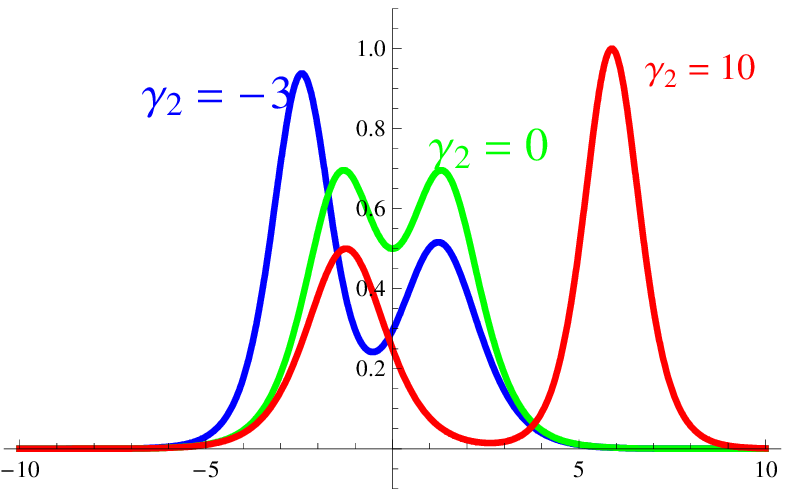}}
\caption{a) Several NTK kink orbits. b) NTK energy densities for
three different values of $\gamma_2$: 1) $\gamma_2=-3$, highest peak
on the left (blue) 2) $\gamma_2=0$, symmetrical peaks (green) 3)
$\gamma_2=10$ highest peak on the right (red).}
\end{figure}

The Hamilton-Jacobi method also provides the energy of the NTK. The
Hamilton characteristic function (the solution of the
Hamilton-Jacobi stationary equation) for zero mechanical energy is:
\[
F(u)+G(v)=(-1)^{{\rm sg}p_u}R \cos\frac{u}{R}+(-1)^{{\rm
sg}p_v}R\cos\frac{v}{R}
\]
From this function we compute the energy of the NTK kinks:
\[
E_{K(\gamma_2)}=2R\left|G(0)-G(f)\right|+2R\left|F(f)-F(\pi-f)\right|
\]
All the NTK kinks have the same energy and satisfy the kink mass sum
rule:
\[
E_{K(\gamma_2)}=2R^2(1+\sigma)=E_{K_2}+E_{K_1}
\]
$K_1$ and $K_2$ are singular kinks that arise at the
$\gamma_2\rightarrow|\infty|$ limit of the NTK kink moduli space.
Their orbits lie on the boundary of the elliptic rectangle and the
$v=0$ axis, see Figure 7. The $K_1$ kink orbit is the straight line
$v_{K_1}(x)=0$. The $K_2$ kink orbit, however, is a three-step
trajectory: the $u=f$, $u=\pi-f$, and $v=\pm f$ edges of the
rectangle.

Observe that all the NTK orbits starting from the North Pole $A$
(South Pole $\bar{A}$) meet at one of the antipodal foci
$\bar{F}_1$-$\bar{F}_2$ (foci $F_1$-$F_2$), which are thus conjugate
points to the Poles. According to the Morse index theorem, these
kinks are unstable, see \cite{ItoT}, \cite{MG}. The $K_2$ orbits
also cross the foci and only the kinks of type $K_1$ are stable.
\begin{figure}[htbp]
\centerline{\includegraphics[height=3cm]{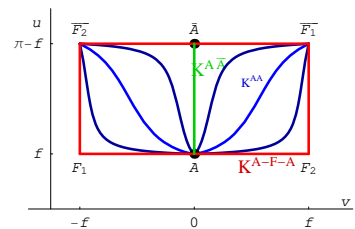}}
\caption{Singular (red and green) and generic (blue) kink orbits in
the elliptic rectangle.}
\end{figure}

\section{Solitary spin  waves}

\noindent The Wess-Zumino action
\[
S_{\rm WZ}=R^2\int \, dtdx \, \sum_{a=1}^3\,
A_a[\Phi(t,x)]\frac{\partial\phi_a}{\partial t}(t,x)
\]
produces the Euler-Lagrange equation:
\[
\frac{\partial A_a}{\partial t}=\sum_{b=1}^3\, \left(\frac{\delta
A_b}{\delta\phi_a}-\frac{\delta
A_a}{\delta\phi_b}\right)\cdot\frac{\partial\phi_b}{\partial
t}=\sum_{b=1}^3\,\sum_{c=1}^3 \,
\varepsilon_{abc}B_c[\Phi]\cdot\frac{\partial\phi_b}{\partial t}
\]
We choose the non-null components of the \lq\lq vector potential" in
the North and South hemispheres of the target space in the form
(using $\varepsilon_{\alpha\beta}=-\varepsilon_{\beta\alpha}$,
$\varepsilon_{12}=1$):
\[
A_\alpha^\pm[\Phi(t,x)]=-\sum_{\beta=1}^2\varepsilon_{\alpha\beta}\frac{\phi_\beta(t,x)}{2R(\phi_3(t,x)\pm
R)} \, \, , \, \, \alpha=1,2
\]
A \lq\lq magnetic monopole field" arises in the target space:
$B_a[\Phi(t,x)]=\frac{\phi_a(t,x)}{R^3}$. The combined
Euler-Lagrange equations for $S_{\rm WZ}+S$  are:
\begin{equation}
\frac{1}{R}\sum_{b=1}^3\, \sum_{c=1}^3 \,
\varepsilon_{abc}\phi_c(t,x)\frac{\partial\phi_b}{\partial
t}(t,x)+\Box\phi_a(t,x)+\frac{\delta V}{\delta\phi_a}(t,x)=0
\label{LL}
\end{equation}

At the long wavelength limit $\omega << \frac{1}{R}$, system
(\ref{LL}) becomes the Landau-Lifhsitz equations for ferromagnetism
with a dispersion relation: $\omega^2(k)=R^2(k^2+1)(k^2+\sigma^2 )$.
The connection between the semi-classical (high-spin) limit of the
Heisenberg model with the quantum non-linear sigma model is well
established \cite{Hal}. Thus, our kinks, which are also static
finite energy solutions of (\ref{LL}), are solitary spin waves in
the low-energy regime of quantum ferromagnets, although the symmetry
is contracted from Lorentzian to Galilean.

\section{Conclusions}

In this letter we have reached the following conclusions about the
kink manifold of the 1D massive non-linear ${\mathbb S}^2$-sigma
model:
\begin{enumerate}

\item If the masses of the pseudo-Goldstone particles are equal,
there exist a ${\mathbb S}^1$-family (fixed the kink center of mass)
of topological kinks degenerated in energy living on all the
half-meridians of the ${\mathbb S}^2$-sphere. When the masses
differ, only two pairs of topological kinks survive, each pair of
kinks having distinct energy.

\item  Even if the masses of the pseudo-Goldstone particles are
different, we have shown that there exist a one-parametric family
(for fixed CM) of non-topological kinks degenerated in energy by
using elliptic coordinates on the ${\mathbb S}^2$-sphere.

\item It is also shown that there is a curious kink mass sum rule
between the non-topological and topological kinks and that only one
of the topological kink pairs is formed by stable kinks.

\item Finally, we have noticed by adding a Wess-Zumino-type term to
the action that our kinks are solitary spin waves in the long
wavelength limit of ferromagnetic materials.

\end{enumerate}

\section{Acknowledgements}

We are grateful to Mariano Santander for enlightening explanations
about elliptic coordinates on the sphere and other Cayley-Klein
two-dimensional spaces.

We thank also to the Spanish Ministerio de Educacion y Ciencia for
partial support under grant FIS2006-09417.

\end{document}